\documentclass[a4paper,aps,preprintnumbers,showpacs,amsmath,amssymb,prl]{revtex4}
\usepackage{mathrsfs}
\usepackage{amsmath}
\usepackage[dvips]{graphicx}
\usepackage{epsfig}
\usepackage[dvips]{graphics,graphicx}
\begin{document}

\title{Quantum noise reduction using a cavity with a Bose Einstein condensate}

\author{Aranya B Bhattacherjee}
\affiliation {Department of Physics, ARSD College, University of Delhi(South Campus), New Delhi-110021, India.}

\begin{abstract}
We study an optomechanical system in which the collective density excitations (Bogoliubov modes) of a Bose Einstein condensate (BEC) is coupled to a cavity field.  We show that the optical force changes the frequency and the damping constant of the collective density excitations of the BEC. We further analyze the occurrence of normal mode splitting (NMS) due to mixing of the fluctuations of the cavity field and the fluctuations of the condensate with finite atomic two-body interaction. The NMS is found to vanish for small values of the two-body interaction. We further show that the density excitations of the condensate can be used to squeeze the output quantum fluctuations of the light beam. This system may serve as an optomechanical control of quantum fluctuations using a Bose Einstein condensate.
\end{abstract}


\maketitle

\section{Introduction}

In recent years mechanical and optical degrees of freedom have become entangled experimentally by underlying mechanism of radiation pressure forces. This field known as cavity optomechanics has played a vital role in the conceptual exploration of the boundaries between classical and quantum mechanical systems. The coupling of mechanical and optical degrees of freedom via radiation pressure has been a subject of early research in the context of laser cooling \cite{hansch,wineland, chu} and gravitational-wave detectors \cite{caves}. Recently there has been a great surge of interest in the application of radiation forces to manipulate the center-of-mass motion of mechanical oscillators covering a huge range of scales from macroscopic mirrors in the Laser Interferometer Gravitational Wave Observatory (LIGO) project \cite{corbitt1, corbitt2} to nano-mechanical cantilevers\cite{hohberger, gigan, arcizet, kleckner, favero, regal}, vibrating microtoroids\cite{carmon, schliesser} membranes\cite{thompson}.  Theoretical work has proposed to use the radiation-pressure coupling for quantum non-demolition measurements of the light field \cite{braginsky}. Recently, coupled dynamics of a movable mirror and atoms trapped in the standing wave light field of a cavity were studied \cite{meiser}. It was shown that the dipole potential in which the atoms move is modified due to the back-action of the atoms and that the position of the atoms can become bistable.

New possibilities for cavity opto-mechanics may emerge by combining the tools of cavity quantum electrodynamics(QED) with those of ultracold gases \cite{brennecke, murch, bhattacherjee09}. Placing an ensemble of atoms inside a high-finesse cavity enhances the atom-light interaction because the atoms collectively couple to the same light mode. The motional degrees of freedom of ultracold atomic gases represent a new source of long-lived coherence affecting light-atom interaction. Nonlinear optics arising from this long-lived coherent motion of ultracold atoms trapped within a high-finesse Fabry-Perot cavity was reported recently \cite{murch}. Strong optical nonlinearities were observed even at low mean photon number of $0.05$. This nonlinearity also gives rise to bistability in the transmitted probe light through the cavity.

It is well known that by using a resonant optical cavity around a Kerr medium, the quantum fluctuations of an incoming light beam can be reduced below the standard quantum-noise limit for a given quadrature component. The quantum optical properties of a mirror coupled via radiation pressure to a cavity field show interesting similarities to an intracavity Kerr-like interaction \cite{fabre, mancini} and can be used to reduce quantum noise of light field reflected by such a cavity.

In this work, we study another kind of Kerr type medium, namely the collective motion of a trapped macroscopic ensemble of ultracold gas coupled to the intensity of the light field inside a cavity which serve as a mechanical oscillator ( analogous to a movable mirror ). First we show how the optical force modifies the frequency and damping constant of the collective density excitations of the BEC and further show the occurrence of normal mode splitting due to mixing of the optical mode and the Bogoliubov mode. Finally we demonstrate for the first time that the density excitations of the condensate can be used to reduce the quantum noise of the cavity field reflected by the BEC. The role of the two-body interaction on the dynamics of the coupled system is also explored. This system may serve as an optomechanical control of quantum fluctuations using a Bose Einstein condensate.

\section{Quantum Langevin equations for the system}

We consider an elongated cigar shaped Bose-Einstein condensate(BEC) of $N$ two-level $^{87} Rb$ atoms in the $|F=1>$ state with mass $m$ and frequency $\omega_{a}$ of the $|F=1>\rightarrow |F'=2>$ transition of the $D_{2}$ line of $^{87} Rb$, strongly interacting with a quantized single standing wave cavity mode of frequency $\omega_{c}$ . The standing wave that forms in the cavity results in an one-dimensional optical lattice potential. The cavity field is also coupled to external fields incident from one side mirror.  It is well known that high-Q optical cavities can significantly isolate the system from its environment, thus strongly reducing decoherence and ensuring that the light field remains quantum-mechanical for the duration of the experiment. We also assume that the induced resonance frequency shift of the cavity is much smaller than the longitudinal mode spacing, so that we restrict the model to a single longitudinal mode.  In order to create an elongated BEC, the frequency of the harmonic trap along the transverse direction should be much larger than one in the axial (along the direction of the optical lattice) direction. The system is also coherently driven by a laser field with frequency $\omega_{p}$ through the cavity mirror with amplitude $\eta$. This system is modeled by the opto-mechanical Hamiltonian $(H_{om})$ in a rotating wave and dipole approximation.

\begin{equation}\label{hom}
H_{om}=\dfrac{p^2}{2m}-\hbar \Delta_{a} \sigma^{+} \sigma^{-} -\hbar \Delta_{c}\hat{a}^{\dagger}\hat{a} - i\hbar g(x)\left[ \sigma^{+}\hat{a}-\sigma^{-}\hat{a}^{\dagger}\right]-i\eta(\hat{a}-\hat{a}^{\dagger})\,
\end{equation}

where $\Delta_{a}=\omega_{p}-\omega_{a}$ and $\Delta_{c}=\omega_{p}-\omega_{c}$ are the large atom-pump and cavity-pump detuning, respectively.  Here $\sigma^{+} , \sigma^{-}$ are the Pauli matrices. The atom-field coupling is written as $g(x)=g_{0} \cos(kx)$. Here $\hat{a}$ is the annihilation operator for a cavity photon. The input laser field populates the intracavity mode which couples to the atoms through the dipole interaction. The field in turn is modified by the back-action of the atoms. The system we are considering is intrinsically open as the cavity field is damped by the photon-leakage through the massive coupling mirror.  Since the detuning $\Delta_{a}$ is large, spontaneous emission is negligible and we can adiabatically eliminate the excited state using the Heisenberg equation of motion $\dot{\sigma^{-}}=\dfrac{i}{\hbar}\left[ H_{om},\sigma^{-}\right] $. This yields the single particle Hamiltonian

\begin{equation}
H_{0}=\dfrac{p^2}{2m}-\hbar \Delta_{c}\hat{a}^{\dagger}\hat{a}+\cos^2(kx)\left[ V_{\text{cl}}({\bf{r}})+\hbar U_{0}\hat{a}^{\dagger} \hat{a}\right] - i\eta(\hat{a}-\hat{a}^{\dagger}).
\end{equation}

The parameter $U_{0}=\dfrac{g_{0}^{2}}{\Delta_{a}}$ is the optical lattice barrier height per photon and represents the atomic backaction on the field . $V_{\text{cl}}({\bf{r}})$ is the external classical potential. Here we will always take $U_{0}>0$. In this case the condensate is attracted to the nodes of the light field and hence the lowest bound state is localized at these positions which leads to a reduced coupling of the condensate to the cavity compared to that for $U_{0}<0$.  Along $x$, the cavity field forms an optical lattice potential of period $\lambda/2$ and depth ($\hbar U_{0}<\hat{a}^{\dagger}\hat{a}>+V_{cl}$). We now write the Hamiltonian in a second quantized form including the two body interaction term.

\begin{eqnarray}
H&=&\int d^3 x \Psi^{\dagger}(\vec{r})H_{0}\Psi(\vec{r})\nonumber \\&+&\dfrac{1}{2}\dfrac{4\pi a_{s}\hbar^{2}}{m}\int d^3 x \Psi^{\dagger}(\vec{r})\Psi^{\dagger}(\vec{r})\Psi(\vec{r})\Psi(\vec{r})\;
\end{eqnarray}

where $\Psi(\vec{r})$ is the field operator for the atoms. Here $a_{s}$ is the two body $s$-wave scattering length. The corresponding opto-mechanical-Bose-Hubbard (OMBH) Hamiltonian can be derived by writing $\Psi(\vec{r})=\sum_{j} \hat{b}_{j} w(\vec{r}-\vec{r}_{j})$, where $w(\vec{r}-\vec{r}_{j})$ is the Wannier function and $\hat{b}_{j}$ is the corresponding annihilation operator for the bosonic atom at the $j^{th}$ site. Retaining only the lowest band with nearest neighbor interaction, we have

\begin{eqnarray}
H &=& E_{0}\sum_{j}\hat{b}_{j}^{\dagger}\hat{b}_{j}+E\sum_{j}\left(\hat{b}_{j+1}^{\dagger}\hat{b}_{j}+\hat{b}_{j+1}\hat{b}_{j}^{\dagger} \right)\nonumber \\&+& (\hbar U_{0}\hat{a}^{\dagger}\hat{a}+V_{cl})\left\lbrace J_{0}\sum_{j}\hat{b}_{j}^{\dagger}\hat{b}_{j}+J \sum_{j}\left(\hat{b}_{j+1}^{\dagger}\hat{b}_{j}+\hat{b}_{j+1}\hat{b}_{j}^{\dagger} \right)\right\rbrace+\dfrac{U}{2}\sum_{j}\hat{b}_{j}^{\dagger}\hat{b}_{j}^{\dagger}\hat{b}_{j}\hat{b}_{j}\nonumber \\&-&\hbar \Delta_{c} \hat{a}^{\dagger}\hat{a}-i\hbar \eta (\hat{a}-\hat{a}^{\dagger})\;
\end{eqnarray}

where

\begin{eqnarray}
U&=&\dfrac{4\pi a_{s}\hbar^{2}}{m}\int d^3 x|w(\vec{r})|^{4}\nonumber \\
E_{0}&=&\int d^3 x w(\vec{r}-\vec{r}_{j})\left\lbrace \left( -\dfrac{\hbar^2 \nabla^{2}}{2m}\right) \right\rbrace w(\vec{r}-\vec{r}_{j})\nonumber \\
E &=&\int d^3 x w(\vec{r}-\vec{r}_{j})\left\lbrace \left( -\dfrac{\hbar^2 \nabla^{2}}{2m}\right) \right\rbrace w(\vec{r}-\vec{r}_{j \pm 1})\nonumber \\
J_{0}&=&\int d^3 x w(\vec{r}-\vec{r}_{j}) \cos^2(kx)w(\vec{r}-\vec{r}_{j})\nonumber \\
J &=&\int d^3 x w(\vec{r}-\vec{r}_{j}) \cos^2(kx)w(\vec{r}-\vec{r}_{j \pm 1}).
\end{eqnarray}

The OMBH Hamiltonian derived above is valid only for weak atom-field nonlinearity \cite{larson}. The nearest neighbor nonlinear interaction terms are usually very small compared to the onsite interaction and are neglected as usual. We now write down the Heisenberg-Langevin equation of motion for the bosonic field operator $\hat{b}_{j}$ and the internal cavity mode $\hat{a}$ as

\begin{eqnarray}\label{lngvn1}
\dot{\hat{b}}_{j}&=&-i(U_{0}\hat{a}^{\dagger} \hat{a}+\dfrac{V_{cl}}{\hbar}) \left\lbrace J_{0}\hat{b}_{j}+J (\hat{b}_{j+1}+\hat{b}_{j-1}) \right\rbrace-\dfrac{iE}{\hbar}\left\lbrace\hat{b}_{j+1}+\hat{b}_{j-1}  \right\rbrace\nonumber \\&-& \dfrac{iU}{\hbar}\hat{b}^{\dagger}_{j} \hat{b}_{j} \hat{b}_{j}-\dfrac{iE_{0}}{\hbar} \hat{b}_{j}-\dfrac{\Gamma_b}{2}\hat{b}_{j}+\sqrt{\Gamma_b/M} \xi_{b}(t)\;
\end{eqnarray}

\begin{eqnarray}\label{lngvn2}
\dot{\hat{a}}&=&-iU_{0}\left\lbrace J_{0}\sum_{j}\hat{b}_{j}^{\dagger}\hat{b}_{j}+J \sum_{j} \left(\hat{b}_{j+1}^{\dagger}\hat{b}_{j}+\hat{b}_{j+1}\hat{b}_{j}^{\dagger} \right)\right\rbrace \hat{a}+\eta\nonumber \\&+&i \left\lbrace \Delta_{c} -\dfrac{\kappa}{2} \right\rbrace \hat{a} +\sqrt{\kappa} \xi_{p}(t)\;
\end{eqnarray}

Here $\kappa$ and $\Gamma_{b}$ characterizes the dissipation of the cavity field and collective density excitations of the BEC resectively. Here, we follow a semi-classical theory by considering \textit{noncommuting} noise operators for the input field, i.e., $\langle\xi_{p}(t)\rangle=0$, $\langle\xi_{p}^{\dagger}(t^{\prime})\xi_{p}(t)\rangle=n_{p}\delta(t^{\prime}-t),\,\langle\xi_{p}(t^{\prime})\xi_{p}^{\dagger}(t)\rangle=\left(n_{p}+1\right)\delta(t^{\prime}-t)$, and a \textit{classical} thermal noise input for the BEC oscillator,
i.e.~$\langle\xi_{b}(t)\rangle=0$, $\langle\xi_{b}^{\dagger}(t^{\prime})\xi_{b}(t)\rangle=\langle\xi_{b}(t^{\prime})\xi_{b}^{\dagger}(t)\rangle=n_{b}\delta(t^{\prime}-t)$, in Eqs.~(\ref{lngvn1},\ref{lngvn2}). The thermal noise input for the BEC is provided by the thermal cloud of atoms and can be considered classical when $k_{B}T>\hbar \omega_{m}$. Here $T$ is the temperature of the thermal reservoir.  The quantities $n_{b}$ and $n_{p}$ are the equilibrium occupation numbers for the mechanical BEC and optical oscillators, respectively. We consider a deep lattice formed by a strong classical potential $V_{\text {cl}}({\bf r})$, so that the overlap between Wannier functions is small. Thus, we can neglect the contribution of tunneling by putting $E=0$ and $J=0$ . Under this approximation, the matter-wave dynamics is not essential for light scattering. In experiments, such a situation can be realized because the time scale of light measurements can be much faster than the time scale of atomic tunneling. One of the well-known advantages of the optical lattices is their extremely high tunability. Thus, tuning the lattice potential, tunneling can be made very slow~\cite{jaksch}.

\section{Dynamics of small fluctuations: Normal Mode Splitting}

Here we show that the coupling of the cavity field fluctuations and the condensate fluctuations (Bogoliubov mode) leads to the splitting of the normal mode into two modes (Normal Mode Splitting(NMS)). The optomechanical NMS however involves driving two parametrically coupled nondegenerate modes out of equilibrium. The NMS does not appear in the steady state spectra but rather manifests itself in the fluctuation spectra of the mirror displacement. To this end, we shift the canonical variables to their steady-state values (i.e.~$\hat{a}\rightarrow {a}_{s}+\hat{a}$ , $\hat{b}_{j}\rightarrow \dfrac{1}{\sqrt{M}}(\sqrt{N}+\hat{b})$) and linearize to obtain the following Heisenberg-Langevin equations:

\begin{equation}\label{H-L0}
\dot{\hat{b}}=-i\left\lbrace \nu +2 U_{eff}\right\rbrace \hat{b}-iU_{eff}\hat{b}^{\dagger}-ig_{c}(\hat{a}+\hat{a}^{\dagger})-\dfrac{\Gamma_b}{2}\hat{b}+\sqrt{\Gamma_b} \xi_{b}(t)\
\end{equation}

\begin{equation}\label{H-L1}
\dot{\hat{a}} =\left(i\Delta_{d}-\frac{\kappa}{2}\right)\hat{a}-ig_{c}(\hat{b}+\hat{b}^{\dagger})+\sqrt{\kappa}\xi_{p}(t),
\end{equation}

Here, $U_{eff}=\dfrac{Un_{0}}{\hbar}$, $g_{c}=U_{0}J_{0}\sqrt{N}|{a}_{s}|$ , $\nu= U_{0}J_{0}|{a}_{s}|^{2}+\dfrac{V_{cl}J_{0}}{\hbar}+\dfrac{E_{0}}{\hbar}$, $\Delta_{d}=\Delta_{c}-U_{0}NJ_{0}$ is the detuning with respect to the renormalized resonance. In deriving the above equation, we have assumed that ${a}_{s}$ the steady state value of $\hat a$ to be real. $N$ is the total number of atoms in $M$ sites. As before, we assume negligible tunneling ($J=E=0$) and hence we drop the site index $j$ from the atomic operators.. We will always assume $\Gamma_{b}\ll\kappa$. We transform to the quadratures: $X_{p}=\hat{a}+\hat{a}^{\dagger}$, $P_{p}=i(\hat{a}^{\dagger}-\hat{a})$, $X_{b}=\hat{b}+\hat{b}^{\dagger}$, $P_{b}=i(\hat{b}^{\dagger}-\hat{b})$. Note that the steady state values can be obtained by putting $\dot{\hat a}=0$ and $\dot{\hat b}=0$ in Eqns.(\ref{lngvn1},\ref{lngvn2}) and solving for $a_{s}$. This yields a cubic equation in $a_{s}$ which has three real solutions for certain values of the parameters. Out of these three real solutions, two are stable which represents bistability. The steady state values of $\hat a$ and $\hat b$ represent points far from the turning points in the bistable systems. The system reaches a steady state only if it is stable and the condition of stability can be obtained by applying the Routh-Hurwitz criterion to Eqns. (\ref{H-L0},\ref{H-L1}).  In the following we will always be in the stable regime. The displacement spectrum of the condensate in Fourier space for $n_{p}=0$ is found as

\begin{equation}\label{displacement_spectra}
S_{x}(\omega)=\frac{\beta_{1}^{2}}{|d(\omega)|^{2}}\left[4 \Gamma_{b}n_{b}+\frac{8 g_{c}^{2} \kappa (\Delta_{d}^{2}+\omega^{2}+\kappa^{2}/4)}{(\Delta_{d}^{2}-\omega^{2}+\kappa^{2}/4)^2+\omega^{2} \kappa^{2}}\right],
\end{equation}

where,

\begin{equation}
|d(\omega)|^{2}=\{\Omega_{eff}^{2}-\omega^{2}\}^{2}+\omega^{2} \Gamma_{eff}^{2},
\end{equation}

and the effective Bogoliubov mechanical frequency ($\Omega_{eff}$) and the effective Bogoliubov mechanical damping ($\Gamma_{eff}$)
\begin{equation}\label{effective frequency}
\Omega_{eff}^{2}=\beta_{1} \beta_{2}+\frac{4 \Delta_{d} g_{c}^{2} \beta_{1} (\Delta_{d}^{2}-\omega^{2}+\kappa^{2}/4)}{(\Delta_{d}^{2}-\omega^{2}+\kappa^{2}/4)^2+\omega^{2} \kappa^{2}}
\end{equation}

and

\begin{equation}\label{effective damping}
\Gamma_{eff}=\Gamma_{b}- \frac{4 \Delta_{d} g_{c}^{2} \beta_{1} \kappa}{(\Delta_{d}^{2}-\omega^{2}+\kappa^{2}/4)^2+\omega^{2} \kappa^{2}}
\end{equation}

Here $\beta_{1}=\nu+U_{eff}$ and $\beta_{2}=\nu+3 U_{eff}$. This spectrum is characterized by a mechanical susceptibility $\chi(\omega)=1/d(\omega)$ of the condensate that is driven by thermal noise ($\propto n_{b}$) and by the quantum fluctuations of the radiation pressure (quantum back-action).

\begin{figure}[t]

\begin{tabular}{cc}
\includegraphics [scale=0.7] {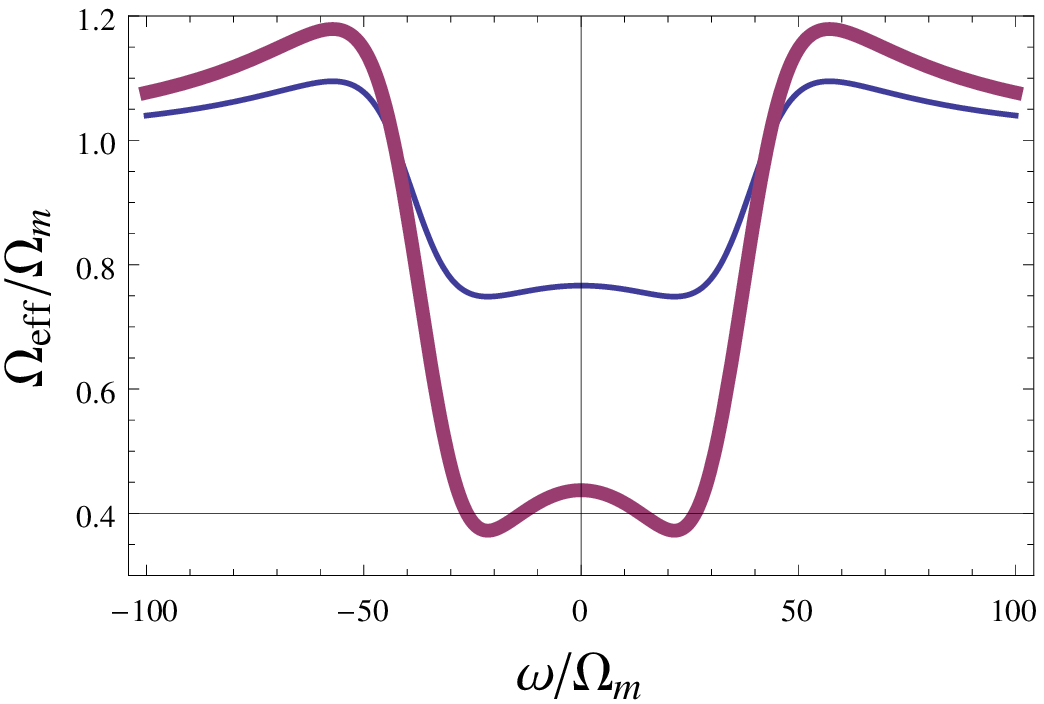}& \includegraphics [scale=0.7] {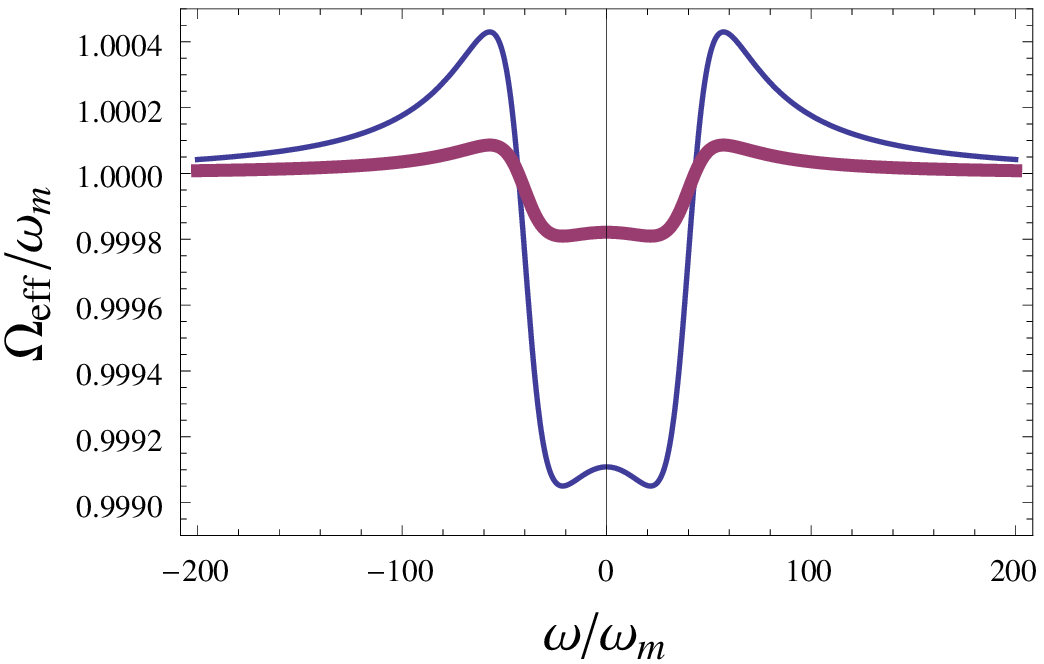}\\
\end{tabular}

\caption{Plot of the normalized effective Bogoliubov mechanical frequency ($\Omega_{eff}/\omega_{m}$, $\omega_{m}=\sqrt{\beta_{1} \beta_{2}}$) of the BEC versus normalized frequency ($\omega/ \omega_{m}$). Parameter values are (Left plot): $n_{b}=10$, $\Gamma_{b}=0.025 \omega_{m}$, $\kappa=32.5 \omega_{m}$, $\Delta_{d}=-40 \omega_{m}$ , $U_{eff}=100 \omega_{m}$ , $\nu= \omega_{m}$ and two values of the atom-photon interaction parameter, $g_{c}=2.5 \omega_{m}$(thin line) and $g_{c}=3.5 \omega_{m}$ (thick line). The right plot shows the normalized effective Bogoliubov mechanical frequency of the BEC versus normalized frequency for two values of the effective two body interaction, $U_{eff}=100 \omega_{m}$(thin line), $U_{eff}=500 \omega_{m}$ (thick line)and $g_{c}=2.5 \omega_{m}$. The other parameters are the same. }

\label{figure1}
\end{figure}

\begin{figure}[t]
\begin{tabular}{cc}

\includegraphics [scale=0.7] {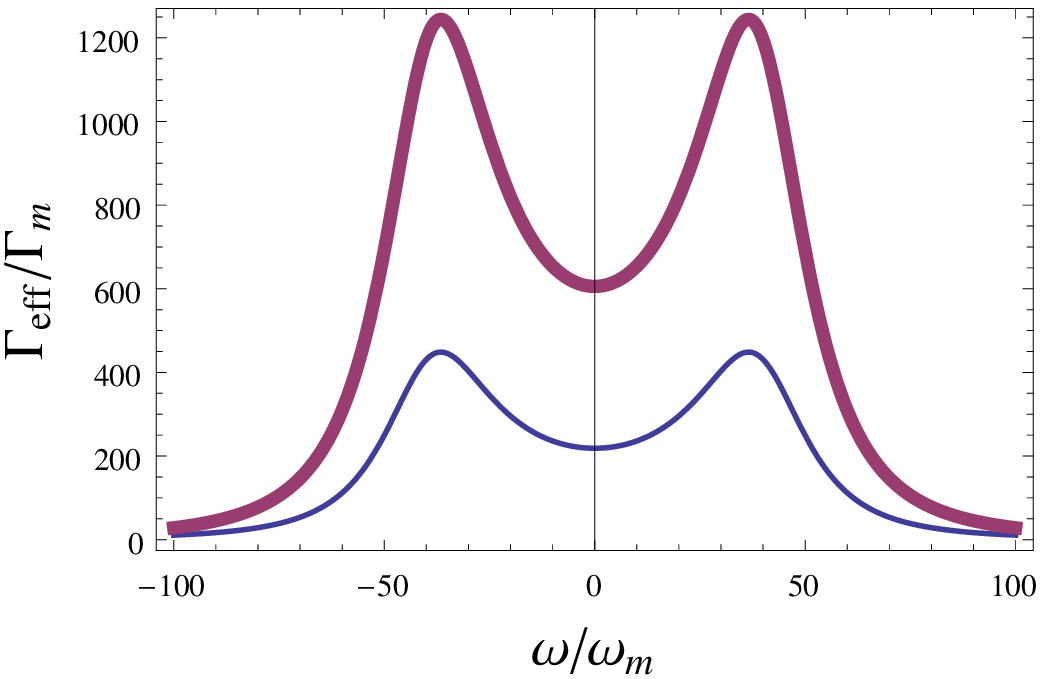} & \includegraphics [scale=0.7] {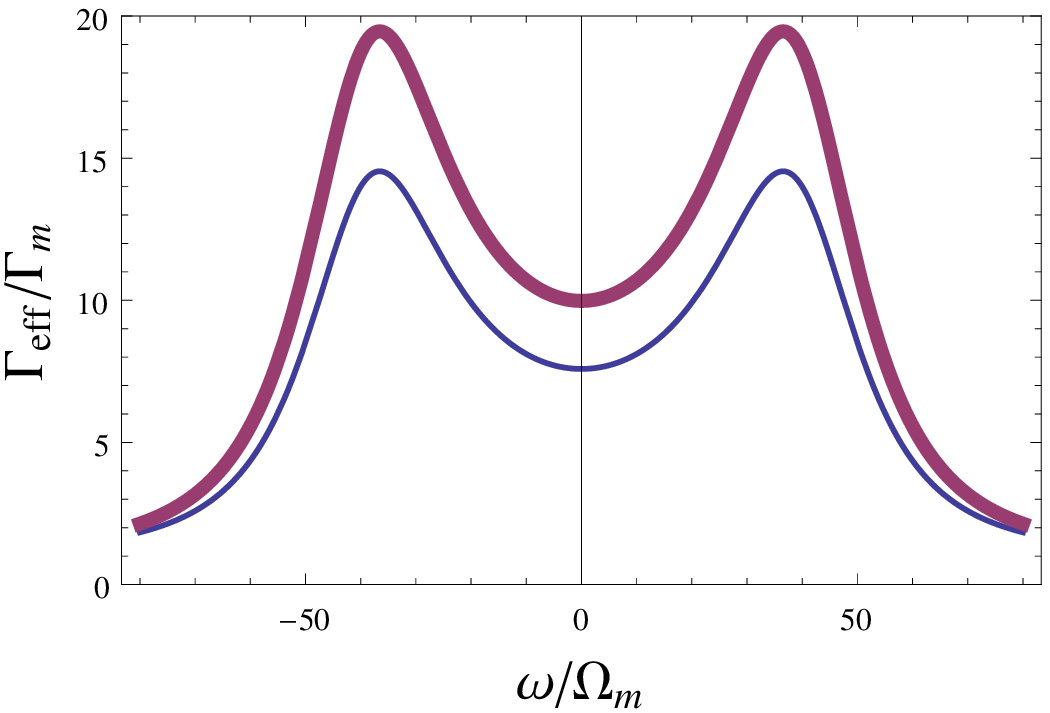}
\end{tabular}
\caption{Plot of the normalized effective Bogoliubov mechanical damping ($\Gamma_{eff}/\omega_{m}$, $\omega_{m}=\sqrt{\beta_{1} \beta_{2}}$) of the BEC versus normalized frequency ($\omega/ \omega_{m}$). Parameter values are (Left plot): $n_{b}=10$, $\Gamma_{b}=0.025 \omega_{m}$, $\kappa=32.5 \omega_{m}$, $\Delta_{d}=-40 \omega_{m}$ , $U_{eff}=100 \omega_{m}$ , $\nu= \omega_{m}$ and two values of the atom-photon interaction parameter, $g_{c}=6.0\omega_{m}$(thin line) and $g_{c}=10 \omega_{m}$ (thick line). The right plot shows the normalized effective Bogoliubov mechanical damping of the BEC versus normalized frequency for two values of the effective two body interaction, $U_{eff}=100 \omega_{m}$(thin line), $U_{eff}=1000 \omega_{m}$ (thick line)and $g_{c}=10 \omega_{m}$. The other parameters are the same. }

\label{figure2}
\end{figure}

The modification of the frequency of the Bogoliubov excitations of the condensate due to the radiation pressure shown by Eq. \ref{effective frequency} is the equivalent of the "optical spring effect" in cavity optomechanical systems with movable mirror. This effect leads to significant frequency shifts in the case of low-frequency oscillations. In Fig.1, we show the plot of the normalized effective Bogoliubov mechanical frequency ($\Omega_{eff}/\omega_{m}$, $\omega_{m}=\sqrt{\beta_{1} \beta_{2}}$) of the BEC versus normalized frequency ($\omega/ \omega_{m}$). Parameter values are (Left plot): $n_{b}=10$, $\Gamma_{b}=0.025 \omega_{m}$, $\kappa=32.5 \omega_{m}$, $\Delta_{d}=-40 \omega_{m}$ , $U_{eff}=100 \omega_{m}$ , $\nu= \omega_{m}$ and two values of the atom-photon interaction parameter, $g_{c}=2.5 \omega_{m}$(thin line) and $g_{c}=3.5 \omega_{m}$ (thick line). The deviation of the Bogoliubov frequency of the condensate from its bare Bogoliubov frequency $\omega_{m}$ increases as the strength of the interaction with the cavity field increases. The right plot shows the normalized effective Bogoliubov mechanical frequency of the BEC versus normalized frequency for two values of the effective two body interaction, $U_{eff}=100 \omega_{m}$(thin line), $U_{eff}=500 \omega_{m}$ (thick line)and $g_{c}=2.5 \omega_{m}$. The other parameters are the same. A higher two body interaction makes the condensate more robust and the Bogoliubov frequency of the condensate does not significantly deviates from $\omega_{m}$.
Figure 2 displays a plot of the normalized effective Bogoliubov mechanical damping ($\Gamma_{eff}/\omega_{m}$, $\omega_{m}=\sqrt{\beta_{1} \beta_{2}}$) of the BEC versus normalized frequency ($\omega/ \omega_{m}$). Parameter values are (Left plot): $n_{b}=10$, $\Gamma_{b}=0.025 \omega_{m}$, $\kappa=32.5 \omega_{m}$, $\Delta_{d}=-40 \omega_{m}$ , $U_{eff}=100 \omega_{m}$ , $\nu= \omega_{m}$ and two values of the atom-photon interaction parameter, $g_{c}=6.0\omega_{m}$(thin line) and $g_{c}=10 \omega_{m}$ (thick line). A stronger coupling with the cavity photons induces a higher atom loss and hence a higher value of the effective damping. This light induced backaction heating and consequent loss of atoms was observed in \cite{murch}. They found that the atom loss rate was enhanced near resonance. The right plot shows the normalized effective Bogoliubov mechanical damping of the BEC versus normalized frequency for two values of the effective two body interaction, $U_{eff}=100 \omega_{m}$(thin line), $U_{eff}=1000 \omega_{m}$ (thick line)and $g_{c}=10 \omega_{m}$. The other parameters are the same. Larger the two body interaction, higher is the damping of the Bogoliubov modes of the BEC. Cooling of the Bogoliubov mode of the BEC by the radiation pressure can be understood in thermodynamical sense. Radiation pressure couples the BEC to the optical cavity mode, which behaves as an effective additional reservoir for the BEC oscillator. As a consequence, the effective temperature of the Bogoliubov mode of the BEC will be intermediate between the initial thermal reservoir temperature and that of the optical reservoir, which is practically zero due to the condition $n_{p}=0$. Therefore one can approach the mechanical ground state of the BEC when the atom-photon coupling rate $g_{c}$ is much larger than the damping rate $\Gamma_{b}$. This explains why significant mechanical cooling of the Bogoliubov mode is obtained when radiation pressure coupling is strong.

Figure 3 shows the normalized plot of the displacement spectrum $S_{x}(\omega)$ of the BEC versus normalized frequency and normalized effective detuning ($\Delta_{d}$) for two values of the atomic two-body interaction, $U_{eff}=150 \times 10^{7} Hz$ (left plot), $U_{eff}=150 \times 10^{5} Hz$(right plot), $\nu=4 \times 10^{4} Hz$, $\Gamma_{b}=735 Hz$ and $g_{c}=\kappa=7.35 \times 10^6 Hz$. In the presence of larger interactions, we observe the usual normal mode spliiting into two modes and we find that if the atom-atom interaction is significantly less, the normal mode splits vanishes (right plot of Figure 3).The NMS is associated with the mixing between the fluctuation of the cavity field around the steady state and the fluctuations of the condensate (Bogoliubov mode) around the mean field. The origin of the fluctuations of the cavity field is the beat of the pump photons with the photons scattered from the condensate atoms. The frequency of the Bogoliubov mode in the low momentum limit is $\approx$ $\sqrt{U_{eff}}$. Hence in the absence of interactions, the Bogoliubov mode is absent and as a result NMS vanishes. In the presence of finite atom-atom interaction, the photon mode and the Bogoliubov mode forms a system of two coupled oscillators. An important point to note is that in order to observe the NMS, the energy exchange between the two modes should take place on a time scale faster than the decoherence of each mode.  Experimentally, Normal mode splitting of a system of large number of atoms coupled to the cavity field has been achieved recently \cite{Klinner06}. It was observed that the NMS was observed only if the coupling between the atoms and the cavity was strong enough (strong cooperative coupling regime). This regime was achieved by increasing the atom numbers. One experimental limitation could be spontaneous emission which leads to momentum diffusion and hence heating of the atomic sample \cite{murch}.

\begin{figure}[t]

\begin{tabular}{cc}
\includegraphics [scale=0.7] {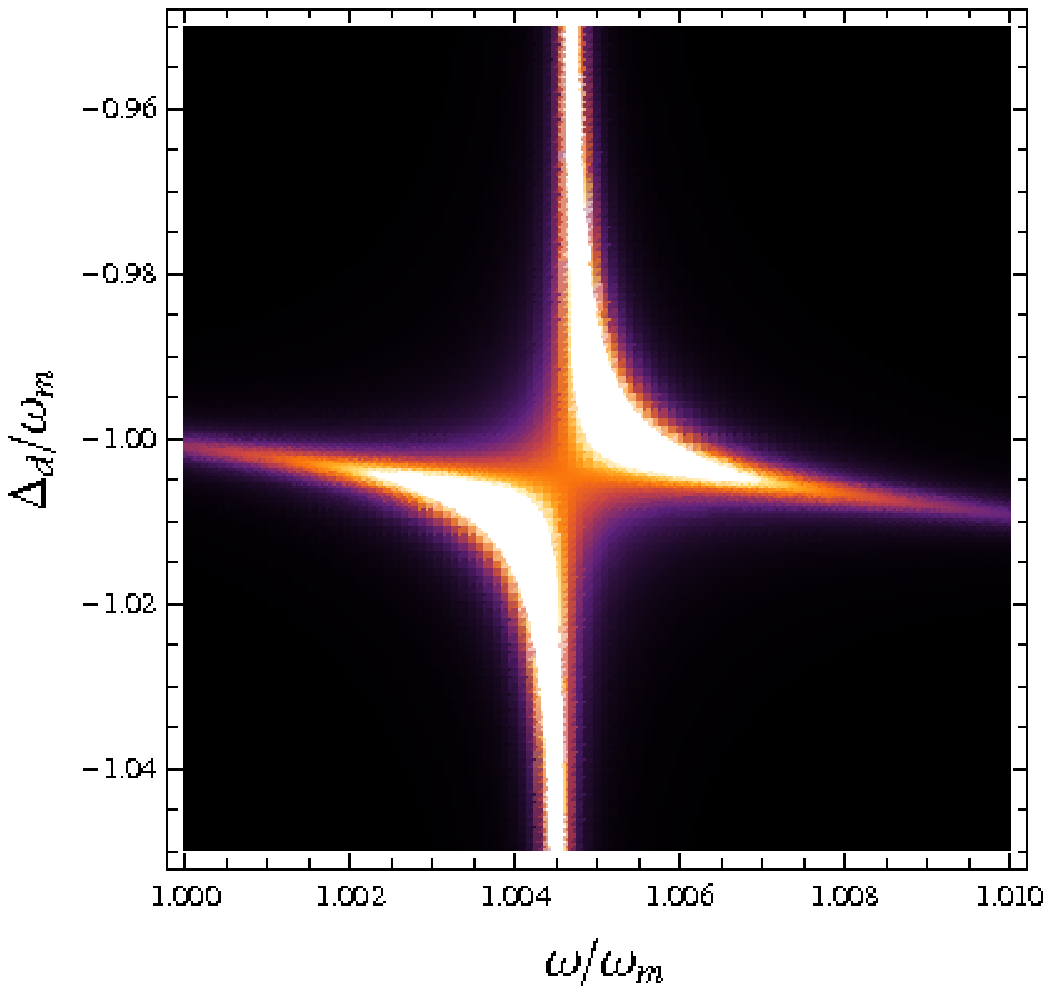} &\includegraphics [scale=0.7] {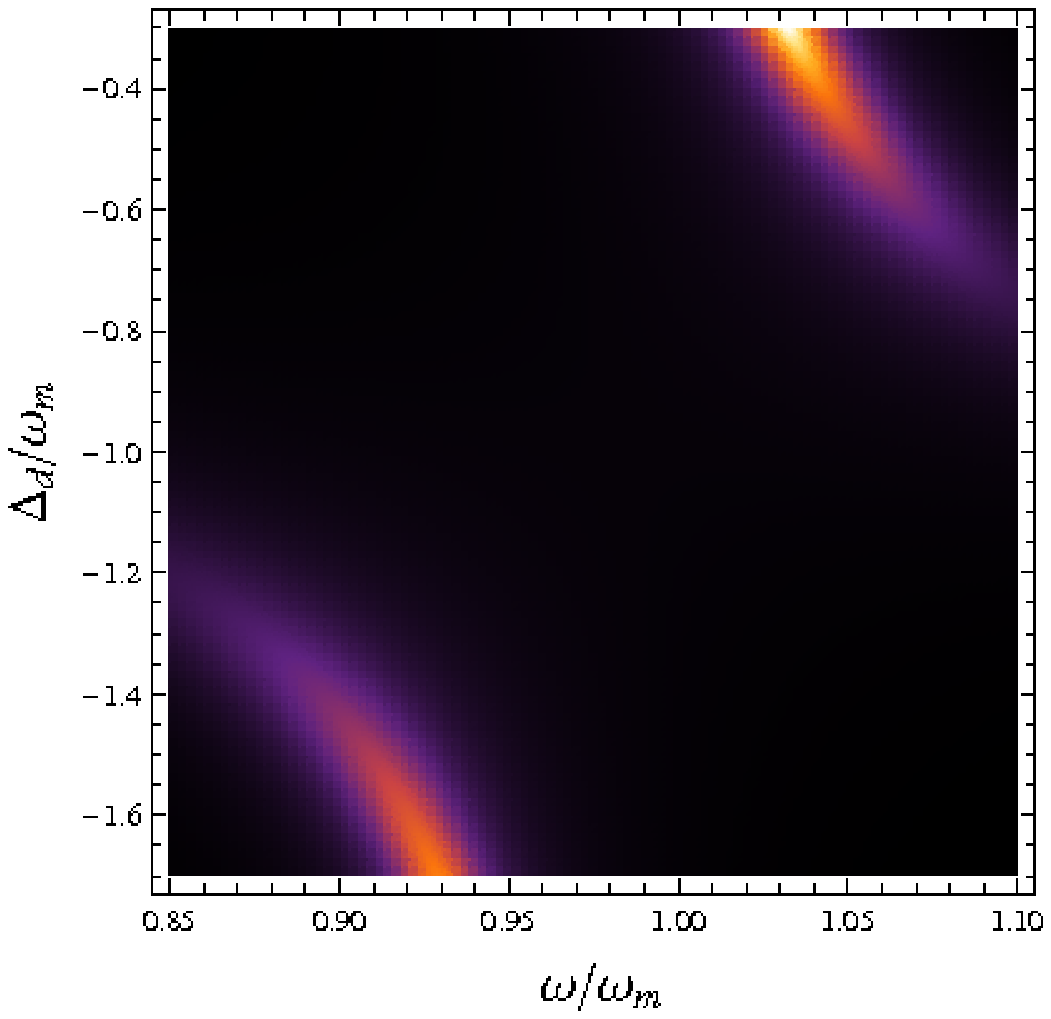}
\end{tabular}

\caption{Normalized plot of the displacement spectrum $S_{x}(\omega)$ of the BEC versus normalized frequency and normalized effective detuning ($\Delta_{d}$) for two values of the atomic two-body interaction, $U_{eff}=150 \times 10^{7} Hz$ (left plot), $U_{eff}=150 \times 10^{5} Hz$(right plot), $\nu=4 \times 10^{4} Hz$, $\Gamma_{b}=735 Hz$ and $g_{c}=\kappa=7.35 \times 10^6 Hz$. Clearly, we see a NMS when $U_{eff}=150 \times 10^{7} Hz$. As the effective interaction decreases, the NMS vanishes. }

\label{figure3}
\end{figure}

\section{Output Intensity Squeezing}
From Eqns. \ref{H-L0} and \ref{H-L1}, we see the affect of the coupling between the cavity mode and the Bogoliubov mode of the BEC. As the intensity of the cavity field is $|a_{s}|^{2}$, there will be an intensity dependent shift of the BEC frequency since $X_{b}$ depends linearly on $\hat a^{\dagger}$, thus introducing a coupling between the fluctuations $\hat a$ and its conjugate $\hat a^{\dagger}$. Thus as a consequence of the dependence of $\hat a$ on $X_{b}$, the fluctuation of the internal cavity field will be squeezed. However $X_{b}$ also depends on $\hat a$, so a further dynamical phase shift and damping is introduced by the coupling of the cavity mode to the BEC.

In frequency space, from the Heisenberg-Langevin equations \ref{H-L0} and \ref{H-L1}, we can find trivially

\begin{equation}\label{squeeze_1}
\{\kappa/2+i (-\Delta_{d}-\omega-|K| \chi(\omega)) \} \hat a(\omega)-i\chi(\omega)K \hat a^{\dagger}(\omega)=\sqrt{\kappa}\xi_{p}(\omega)+ \frac{ig_{c} \sqrt{\Gamma_{b}}\chi(\omega)\xi_{m}(\omega) }{\omega_{m}^{2}},
\end{equation}

where, $|K|=2 \beta_{1} |g_{c}|^{2}/\omega_{m}^{2}$ and here we have introduced the dimensionless dynamical response factor of the BEC:

\begin{equation}
\chi(\omega)=\frac{\omega_{m}^{2}}{(\omega_{m}^{2}-\omega^{2})-i \Gamma_{b} \omega}=\chi_{1}(\omega)+i \chi_{2}(\omega),
\end{equation}

with $\chi^{*}(\omega)=\chi(- \omega)$.

The input-output theory gives the following relation among the incoming field ($\xi_{p}(\omega)$), internal field ($\hat a$) and output field ($\hat a_{out}$) as a consequence of boundary condition at the fixed mirror surface

\begin{equation}
\hat a_{out}(\omega)+ \xi_{p}(\omega)=\sqrt{\kappa} \hat a(\omega),
\end{equation}

Using Eq. \ref{squeeze_1} and its conjugate, we write

\begin{equation}
\hat a_{out}(\omega)=\zeta (\omega) \xi_{p}(\omega)+\eta(\omega) \xi_{p}^{\dagger}(\omega)+\varsigma(\omega) \xi_{m}(\omega),
\end{equation}

and $\hat a_{out}^{\dagger}(\omega)=[\hat a_{out}(-\omega)]^{\dagger}$, where

\begin{equation}
\zeta(\omega)=\frac{\omega^{2}+[\kappa/2+i \Delta_{d}][\kappa/2-2 \chi_{2}(\omega)|K|+i (2 \chi_{1}(\omega)|K|+\Delta_{d})]}{\Delta(\omega)},
\end{equation}

\begin{equation}
\eta(\omega)=\frac{i \kappa \chi(\omega)K}{\Delta(\omega)}=-\eta^{*}(-\omega),
\end{equation}

\begin{equation}
\varsigma(\omega)=\frac{i g_{c} \chi(\omega) \sqrt{\kappa \Gamma_{b}}[\kappa/2-i(\omega-\Delta_{d})]}{\omega_{m}^{2} \Delta(\omega)}
\end{equation}

The output intensity spectrum is $S_{I}(\omega)$ is defined as

\begin{equation}
S_{I}(\omega)=\frac{1}{|\alpha_{out}|^{2}} \int d \omega'<\delta I_{out}(\omega) \delta I_{out}(\omega')>
\end{equation}

and

\begin{equation}
\delta I_{out}(\omega)=\alpha_{out}^{*} \hat a_{out}(\omega)+\alpha_{out} \hat a_{out}^{\dagger}(\omega),
\end{equation}

where, $\alpha_{out}=<\hat a_{out}(\omega)>$ and $\alpha_{in}=<\xi_{p}(\omega)>$. As a consequence of the boundary conditions at the fixed mirrors, $\alpha_{out}=\sqrt{\kappa}a_{s}-\alpha_{in}$. This yield the output intensity spectrum for $n_{p}=0$ as

\begin{equation}\label{squeezing}
S_{I}(\omega)=1+\frac{4 \kappa \omega \Delta_{d} K}{(\kappa^{2}/4+\Delta_{d}^{2})|\Delta(\omega)|^{2}}\{\frac{\omega \Delta_{d}|\chi(\omega)|^{2} n_{b}}{Q}+\chi_{2}(\omega)[\kappa^{2}/4+\Delta_{d}^{2}-\omega \Delta_{d}]+1/2 \omega \kappa \chi_{1}(\omega) \}
\end{equation}

Here, $Q=\omega_{m}/\Gamma_{b}$ is the mechanical quality factor of the condensate. We see that the thermal contribution destroys the squeezing.

\begin{figure}[t]

\begin{tabular}{cc}
\includegraphics [scale=0.7] {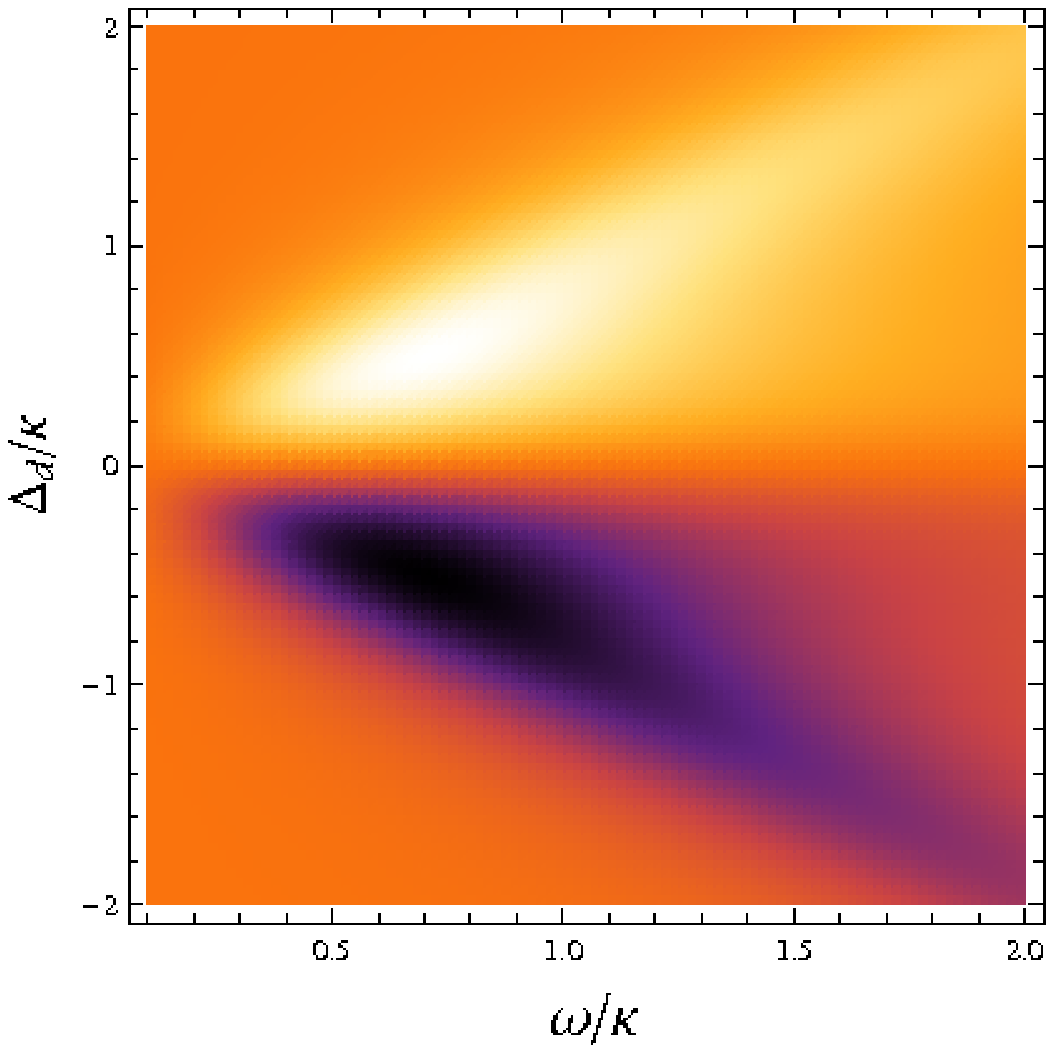} &\includegraphics [scale=0.7] {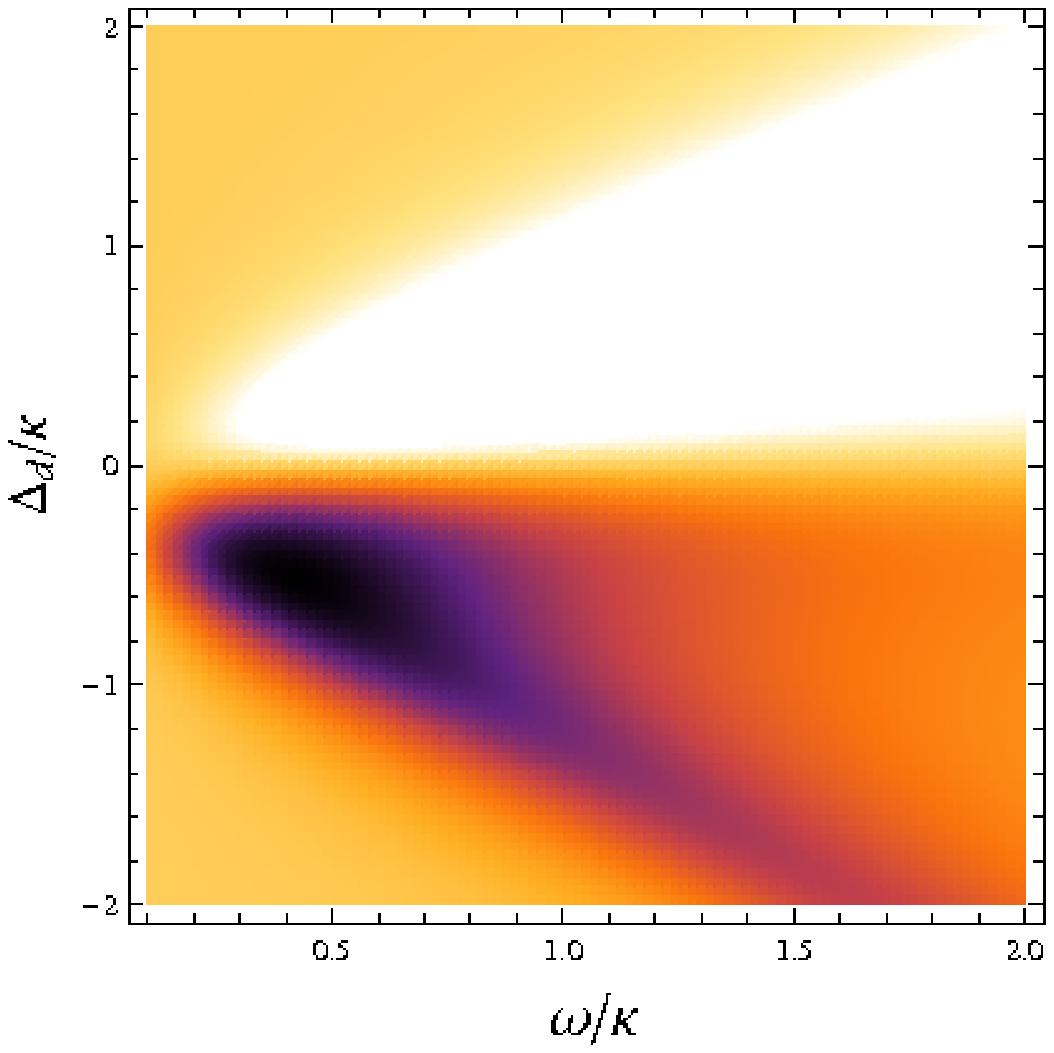}
\end{tabular}

\caption{Normalized intensity squeezing spectrum versus normalized detuning ($\Delta_{d}/ \kappa$) and normalized frequency ($\omega/ \kappa$). The parameters used for the left plot are $K=0.00327 \kappa$, $g_{c}=\kappa$, $\omega_{m}=353.47 \kappa$, $\Gamma_{b}=0.0001 \kappa$ and $n_{m}=10$. For the right plot, $\omega_{m}=3.54 \kappa$, $K=0.326 \kappa$, $g_{c}=\kappa$, $\Gamma_{b}=0.0001 \kappa$ and $n_{m}=10$. In the $(\omega,\Delta_{d})$ parameter space,squeezing of the output light intensity ($S_{I}(\omega)<1$) is indicated by the dark regions.  }

\end{figure}

Figure 4 shows the normalized intensity squeezing spectrum versus normalized detuning ($\Delta_{d}/ \kappa$) and normalized frequency ($\omega/ \kappa$). The parameters used for the left plot are $K=0.00327 \kappa$, $g_{c}=\kappa$, $\omega_{m}=353.47 \kappa$, $\Gamma_{b}=0.0001 \kappa$ and $n_{m}=10$. For the right plot, $\omega_{m}=3.54 \kappa$ (the lower value of $\omega_{m}$ is obtained by lowering the atom-atom interaction $U_{eff}$ ), $K=0.326 \kappa$, $g_{c}=\kappa$, $\Gamma_{b}=0.0001 \kappa$ and $n_{m}=10$. In the $(\omega,\Delta_{d})$ parameter space,squeezing of the output light intensity ($S_{I}(\omega)<1$) is indicated by the dark regions. Higher the squeezing, darker the region. Clearly, we find that the squeezing region in the parameter space $(\omega,\Delta_{d})$ decreases with decreasing two body atom-atom interaction. This is evident as we go from the left plot to the right plot. Significant squeezing is observed for $\Delta_{d}<0$ and $\omega<\kappa$. These results can be easily explained from Eqns. \ref{squeezing}. As mentioned before thermal contribution destroys the squeezing. In order to ensure that the thermal noise contribution does not influence the squeezing $S_{I}(\omega)$, the mechanical quality factor of the condensate has to be large ($\omega_{m}>>\Gamma_{b}$). Thus for the left plot of Fig. 4, the mechanical quality factor $Q=3.54 \times 10^{6}$ and while for the right plot $Q=3.54 \times 10^{4}$. This implies that the thermal noise contribution is enhanced for the right plot and as a consequence the spectrum $S_{I}(\omega)$ exhibits less squeezing in the parameter space $({\omega, \Delta_{d}})$. Further in order to observe enhanced squeezing, the Bogoliubov mechanical frequency of the BEC $\omega_{m}$ should be larger than the cavity linewidth $\kappa$. This can be achieved by increasing the atom-atom interaction $U_{eff}$. If $\omega_{m}<\kappa$, the cavity photons do not see any coherently variable collective position of the BEC (large amplitude of the Bogoliubov mode) but only small fluctuations of the collective position. This can also be interpreted in terms of the coherence length (the coherence length is inversely proportional to $\sqrt{U_{eff}}$). There are two length scales, the coherence length and the spatial scale of variation of the density. At low values of the interaction, the coherence length of the condensate is large and the quantum pressure term dominates the usual pressure term and hence the spatial variations of the density occurs on a length scale less than the coherence length. Under this circumstance, atoms behave as almost free particles. On the other hand, when the interactions are large, the coherence length decreases and the spatial scale of variation of the density become large compared to the coherence length and hence the atoms move collectively. We will then require $\omega_{m}>\kappa$. This implies that when $U_{eff}$ decreases, $\omega_{m}$ also decreases and approaches $\kappa$ and hence a reduction in squeezing in the $({\omega, \Delta_{d}})$ space is observed.

To demonstrate that the dynamics investigated here are within experimental reach, we discuss the experimental parameters from \cite{murch,brennecke}: A BEC of typically $10^{5}$ $^{87}Rb$ atoms is coupled to the light field of an optical ultra high-finesse Fabry-Perot cavity. The atom-field coupling $g_{0}=2 \pi \times 10.9 Mhz$ \cite{brennecke} ( $2 \pi \times 14.4 $ \cite{murch}) is greater than the decay rate of the intracavity field $\kappa=2 \pi \times 1.3 Mhz$ \cite{brennecke} ($2 \pi \times 0.66 Mhz$ \cite{murch}).  Typically atom-pump detuning is $2 \pi \times 32 Ghz$. The rate $\Gamma_{b}$ at which atoms are coupled out of the BEC is about $r \pi \times 7.5 \times 10^{-3} Hz$ \cite{brennecke}. The kinetic energy and potential energy contribution $\nu$ is about $35 kHz$ \cite{brennecke}($49 kHz$ \cite{murch}).The energy of the cavity mode decreases due to the photon loss through the cavity mirrors, which leads to a reduced atom-field coupling. Photon loss can be minimized by using high-Q cavities. Our proposed detection scheme relies crucially on the fact that coherent dynamics dominate over the losses. It is important that the characteristic time-scales of coherent dynamics are significantly faster than those associated with losses.

\section{Conclusions}

In summary we have analyzed a novel scheme of cavity-opto-mechanics with ultracold atoms. We showed that due to the optical force experienced by the BEC in the cavity, the damping rate and the frequency of the Bogoliubov mode of the condensate changes. In the presence of atom-atom interactions, the cavity field fluctuations and the condensate fluctuations (Bogoliubov mode) leads to the splitting of the normal mode into two modes (Normal Mode Splitting).  The system described here shows a complex interplay between distinctly two systems namely,  optical micro-cavity and the gas of ultracold atoms. We found that using a BEC with high mechanical quality factor squeezing of the intensity of the output light field is obtained at low frequency. This scheme may lead to a possible realization of a quantum device to tailor quantum fluctuations of output light using cavity quantum electrodynamics and BEC technology.

\end{document}